\documentclass[lettersize,journal]{IEEEtran}
\usepackage{array}
\usepackage{textcomp}
\usepackage{stfloats}
\usepackage{url}
\usepackage{verbatim}
\usepackage{multirow}
\usepackage[table,xcdraw]{xcolor}
\usepackage{bbm}
\usepackage{cite}
\usepackage{amsmath,amssymb,amsfonts}
\usepackage{algorithmic}
\usepackage{subcaption}
\usepackage{graphicx}
\usepackage{textcomp}
\usepackage{xcolor}
\usepackage{multirow}
\usepackage[ruled,linesnumbered]{algorithm2e}
\usepackage{comment}
\usepackage{float} 
\usepackage{threeparttable}

\DeclareMathAlphabet{\mathbbold}{U}{bbold}{m}{n}

\def\BibTeX{{\rm B\kern-.05em{\sc i\kern-.025em b}\kern-.08em
    T\kern-.1667em\lower.7ex\hbox{E}\kern-.125emX}}
\usepackage{balance}

\begin{document}
\title{\huge{Wireless Resource Allocation with Collaborative Distributed and Centralized DRL under Control Channel Attacks}}
\author{
	Ke Wang, Wanchun Liu, and Teng Joon Lim\\ School of Electrical and Computer Engineering, The University of Sydney, Australia\\ Emails: ke.wang@sydney.edu.au, wanchun.liu@sydney.edu.au, tj.lim@sydney.edu.au.
\vspace{-1.0cm}
}


\maketitle

\begin{abstract}
    In this paper, we consider a wireless resource allocation problem in a cyber-physical system (CPS) where the control channel, carrying resource allocation commands, is subjected to denial-of-service (DoS) attacks. We propose a novel concept of collaborative distributed and centralized (CDC) resource allocation to effectively mitigate the impact of these attacks. To optimize the CDC resource allocation policy, we develop a new CDC-deep reinforcement learning (DRL) algorithm, whereas existing DRL frameworks only formulate either centralized or distributed decision-making problems. Simulation results demonstrate that the CDC-DRL algorithm significantly outperforms state-of-the-art DRL benchmarks, showcasing its ability to address resource allocation problems in large-scale CPSs under control channel attacks. 
\end{abstract}


\section{Introduction}
Cyber-physical systems (CPS) seamlessly integrate physical operations with computational and communication capabilities, enabling advanced control, decision-making, and sensing. As integral components of industrial monitoring systems, wireless sensors and remote estimators face significant network security challenges. The susceptibility of wireless channels to cyberattacks necessitates the development of robust and secure estimation and control strategies to enhance the performance, resilience, and security of CPS \cite{Ding2021Secure}.

Dynamic wireless resource allocation in CPS is frequently modeled as a Markov decision process (MDP)~\cite{LEONG2020Deep, Pang2023DRL, Dai2020Distributed, Wang2024Deep}. As device networks expand, the complexity of MDPs increases correspondingly. Deep reinforcement learning (DRL), which employs neural networks as function approximators, has become a pivotal technique for addressing large-scale MDP challenges within CPS. { DRL can be implemented in centralized or distributed frameworks, each with unique benefits and limitations. Centralized DRL, with access to complete information, improves coordination and optimization but involves higher computational costs and security risks. Distributed DRL, on the other hand, enhances scalability by breaking the problem into independent sub-problems; however, limited access to global information can lead to suboptimal decisions.}
Recent studies have applied DRL to various aspects of remote estimation, adopting both centralized and distributed strategies. Centralized DRL has been used to optimize sensor scheduling~\cite{LEONG2020Deep} and power allocation for remote estimation~\cite{Pang2023DRL}. Distributed DRL has addressed sensor scheduling under data transmission channel denial-of-service (DoS) attacks~\cite{Dai2020Distributed}. A centralized DRL framework has been proposed in ~\cite{Wang2024Deep}, enhancing system resilience against DoS attacks on data channels.

{In addition to data channel attacks, broadcast control channel attacks—where malicious interference targets the channels that distribute essential network management commands to multiple devices simultaneously—pose a significant threat to the stability and security of CPS. By disrupting these broadcasted control signals, attackers can prevent crucial coordination among network components, undermining system integrity. Chan \emph{et al.} addressed broadcast control channel jamming through random mapping with cryptographic functions and deterministic key assignment schemes~\cite{Chan2007Broadcast}. Tague \emph{et al.} expanded on this by introducing a framework that uses random cryptographic key assignments to obscure the locations of these control channels~\cite{Tague2009Mitigation}. Lo \emph{et al.} presented a multi-agent DRL approach that dynamically allocates control channels in response to both primary user activity and jamming attempts~\cite{Lo2012Multiagent}. While these studies contribute to maintaining control channel functionality under attack, the challenge of optimal wireless resource allocation in the presence of broadcast control channel attacks remains largely unexplored. Solely relying on centralized or distributed decision-making may yield suboptimal solutions in addressing this issue.}

Our main contributions in this work are as follows: 

{
$\bullet$ We formulate a novel wireless resource allocation problem for a multi-sensor remote state estimation system facing broadcast control channel DoS attacks. This formulation necessitates both centralized and distributed strategies to effectively mitigate these attacks, addressing a critical challenge previously unexplored in the literature.

$\bullet$ We propose a novel DRL framework, referred to as Collaborative Distributed and Centralized DRL (CDC-DRL), for the collaborative distributed-centralized power allocation problem, which has not been explored in the literature.}

$\bullet$ We demonstrate that CDC-DRL outperforms state-of-the-art centralized and distributed DRL benchmarks, reducing the overall remote estimation error on average by 13.0\% and 52.6\% in small and large scale systems, respectively.

\section{System Model}
\begin{figure}[h]
    \centering
    \includegraphics[width=7cm]{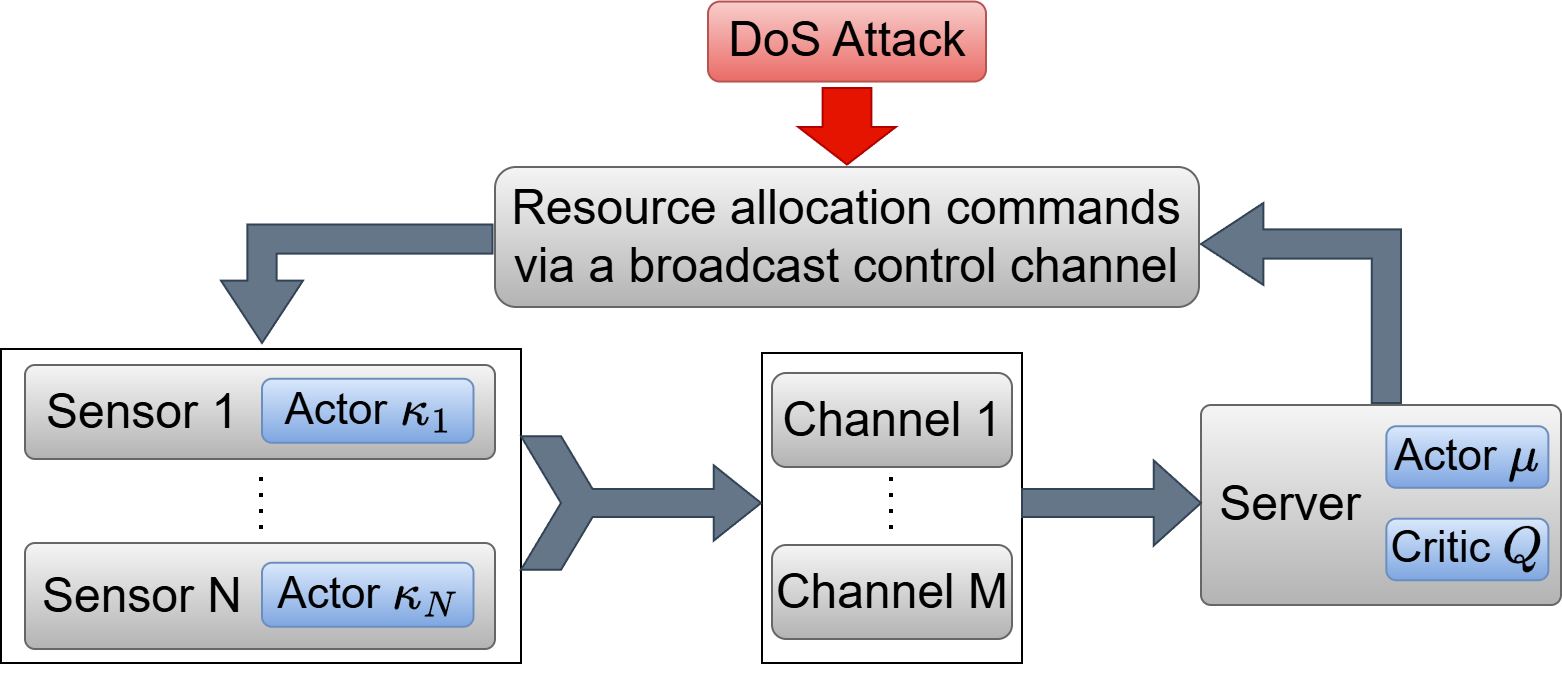}
    \vspace{-0.16cm}
    \captionsetup{font=footnotesize}    
    \caption{Multi-sensor remote state estimation under broadcast control channel DoS attack. Actor and critic modules (in blue) will be explained in Section~\ref{sec:DRL}.}
    \label{fig:system}
    \vspace{-0.5cm}
\end{figure}

We consider a remote estimation system consisting of $N$ sensors and a server (i.e., a remote estimator), as illustrated in Fig.~\ref{fig:system}. The system includes $M$ data channels for sensor information transmission to the server and a broadcast control channel for transmitting resource allocation commands to the sensors.
\vspace{-0.2cm}
\subsection{Remote State Estimation with Smart Sensors}
The physical process measured by sensor $n$ is modeled as a discrete-time linear system~\cite{LEONG2020Deep}:
\begin{equation}
    \begin{aligned}
        & x_{n, k+1} = A_{n} x_{n, k} + \omega _{n, k}, \\
        & y_{n, k} = C_{n} x_{n, k} + \nu_{n, k}, 
    \end{aligned}
\end{equation}
where non-negative integer $k \in \mathbb{Z}^{+}$ is the time index. $x_{n, k}$ is the system state. $A_n$ is the state transition matrix. $y_{n, k}$ is measurement of sensor $n$. $C_n$ is the measurement matrix. $\omega_{n, k}$ and $\nu_{n, k}$ represent i.i.d. Gaussian process and measurement noise with covariance matrix $\Omega_n$ and $\Theta_n$, respectively. Instead of sending the raw measurement $y_{n,k}$ directly to the server, each sensor with computational capability (i.e., a ``smart" sensor) pre-processes its measurement data by running a local Kalman filter and sends its local state estimates $\widehat{x}_{n, k}^{s}$. As we focus on the remote estimation scenario, we assume that each local Kalman filter is stable and working at its steady state with the estimation error covariance matrix $\overline{P_{n}} \triangleq \mathbb{E}[(x_{n,k}-\widehat{x}^s_{n,k})(x_{n,k}-\widehat{x}^s_{n,k})^\top], \forall k \in \mathbb{Z}^{+}$~\cite{Pang2023DRL, Dai2020Distributed, Wang2024Deep}. 

Due to limited wireless communication resources and potential packet errors, the server may not receive all local sensor estimates in each time slot and thus employs a Minimum Mean-Square Error (MMSE) estimator for each process~\cite{Schenato2008Optimal}. We define the binary variable $\gamma_{n, k} \in \{0,1\}$ such that successful transmission of sensor $n$'s packet is indicated by $\gamma_{n, k} = 1$.\footnotemark 
\renewcommand{\thefootnote}{1}
\footnotetext{Successful data reception is acknowledged to the sensor with a 1-bit feedback signal, a standard approach in systems like 5G NR and LTE. Due to its minimal data load, this feedback channel is assumed to be attack-free.}The remote state estimate and the error covariance matrix are given by 
\begin{equation}
\begin{aligned}
\widehat{x}_{n, k} & = \begin{cases}\widehat{x}_{n, k}^{s} & \text{if} \ \gamma_{n, k} = 1\\A_{n} \widehat{x}_{n, k-1}& \text{otherwise, }\end{cases} \\
P_{n, k} & = \begin{cases} \overline{P_{n}} & \text{if} \ \gamma_{n, k} = 1\\ h_{n}(P_{n, k-1})& \text{otherwise, }\end{cases}
\end{aligned}
\end{equation}
where \begin{math}h_{n}(X) \triangleq A_{n}X A_{n}^{\top} + W_{n}\end{math} is the Lyapunov operator.
Define the age of information (AoI) at sensor $n$ at time $k$ as 
\begin{equation}\label{eq:tau_update}
    \tau_{n, k} \triangleq \min \{ t \geq 0: \gamma_{n, k-t} = 1 \} = \begin{cases}
        0 & \text{if} \ \gamma_{n, k} = 1\\
        \tau_{n, k-1} \! + \! 1 & \text{otherwise,}
    \end{cases} 
\end{equation}
allowing $P_{n, k}$ to be rewritten as
\begin{equation}
    P_{n, k} = h_n^{\tau_{n, k}} (\overline{P_{n}}).
\end{equation}
Now, we define the expected estimation error at time $k$ for measuring the remote estimation performance as 
\begin{equation}
\label{eq:ek}
c_k \triangleq \sum_{n=1}^N \mathrm{tr} P_{n,k},
\end{equation}
where $\mathrm{tr}$ is the trace operator. The term $c_k$ captures the estimation MSE across all sensors and depends on the sensors' AoI, $\tau_{1,k},\dots,\tau_{N,k}$.

\subsection{NOMA-based Data Transmission Channel}\label{sec:channel}
We consider $M$ orthogonal channels (e.g., subcarriers) for uplink sensor data transmissions. 
To model the time-varying and correlated nature of wireless fading channels, we use a time-homogeneous, ergodic Markov block-fading channel model with $\psi$ quantization levels~\cite{Sadeghi2008Finite}. Each channel's power gain is constant within a packet transmission time slot but may vary between packets~\cite{liu2020remote}. Let $\mathbf{g}_{n, k} \triangleq [g^{n}_{1, k}, \ldots, g^{n}_{M, k}]^\top$ denote the channel power gain from sensor $n$ to the server across $M$ channels, where $g^{n}_{m, k} \in \mathbb{G} \triangleq \{h_{1}, \ldots, h_{\psi}\}$ represents the discrete state of the channel power gain on channel $m$. The sensor's channel state vector $\mathbf{g}_{n, k} \in \{\mathbf{h}_1, \ldots, \mathbf{h}_{\bar{\psi}} \}$ is modeled as a multi-state Markov chain with $\bar{\psi} = \psi^M$ states. Assuming independent Markov channel states for different sensors due to their unique locations and varying radio propagation conditions, the channel state transition matrix $\mathbf{M}_{n} \in \mathbb{R}^{\bar{\psi} \times \bar{\psi}}$ for sensor $n$ has the $i, j$-th element representing the transition probability from state $\mathbf{h}_{i}$ to $\mathbf{h}_{j}$.

We adopt an advanced Non-Orthogonal Multi-Access (NOMA) scheme based on~\cite{Pang2023DRL}. Each channel can accommodate multiple concurrent sensors, and each sensor can send its packet across one or more of the $M$ orthogonal channels given a total power budget $E^{s}$.
The transmission power of sensor $n$ in all $M$ channels at time slot $k$ is captured in the vector
\begin{equation}
\label{eq:overall_action}
\mathbf{a}_{n, k} \triangleq [E^{n}_{1, k}, \ldots, E^{n}_{M, k}]^{\top},
\end{equation}
where $\sum_{m = 1}^{M} E^{n}_{m, k} \leq E^{s}$. The aggregate transmission power of all sensors is represented by $\mathbf{a}_{k} \triangleq [\mathbf{a}_{1, k}, \dots, \mathbf{a}_{N, k}]$.

The server employs an \(N\)-round Interference Rejection Combining (IRC) and Successive Interference Cancellation (SIC) algorithm to recover each packet from the \(N\) sensors from the superimposed signals across all \(M\) channels. In each iteration, the server computes the Signal-to-Interference-plus-Noise Ratio (SINR) for the undecoded sensors using IRC and executes SIC to detect the packet with the highest SINR. The process terminates when either a packet detection error occurs or all packets are detected successfully.

Assuming each sensor packet has a block length of \(L\) and data rate \(R\), the packet drop probabilities for each sensor can be accurately approximated based on the detailed SINR calculation and finite-blocklength information theory~\cite{Yury2010Channel}, as shown in~\cite{Pang2023DRL}. Further specifics are omitted in this paper due to space limitations.

\subsection{DoS Attacker at Broadcast Control Channel}
The server generates transmit power allocation commands for all sensors, packages this information into a single packet, and sends it over the broadcast control channel.
A DoS attacker with Markovian behavior targets the broadcast control channel by transmitting a wireless jamming signal. Let the attacker's energy budget be $E^a$ and its actual energy expended at time $k$ be $E^a_k$. To minimize the likelihood of detection, the attacker intermittently alternates between attack ($E^a_k = E^a$) and silence ($E^a_k = 0$). The 2-by-2 attacker's action transition matrix is denoted by $T^a$.

Let $g^{d}_{n, k}$ and $g^{a}_{n, k} \in \mathbb{G}$ denote the control channel gain and the jamming channel (i.e., from attacker to sensor) gain for sensor $n$, respectively. The signal-to-interference-plus-noise ratio (SINR) on the broadcast control channel for sensor $n$ is defined as
\begin{equation}
    \rho^{d}_{n, k} = \frac{g^{d}_{n, k} E^{d}}{g^{a}_{n, k} E^a_{k} + \sigma^2},
\end{equation}
where $\sigma^2$ is the sensor's receiver noise variance and $E^{d}$ is the server's transmit power on the control channel. The packet drop probability of the control channel can be approximated as for the data channel. 

\subsection{Collaborative Distributed and Centralized Power Allocation}
We propose a collaborative distributed and centralized power allocation scheme to mitigate the impact of broadcast control channel attacks. Let $\mathbf{a}^c_{k} \triangleq [\mathbf{a}^c_{1, k}, \dots, \mathbf{a}^c_{N, k}]$ denote the centralized power allocation action generated by the server. The distributed power allocation action generated by sensor $n$ is denoted as $\mathbf{a}^d_{n, k}$, and $\mathbf{a}^d_{k} \triangleq [\mathbf{a}^d_{1, k}, \dots, \mathbf{a}^d_{N, k}]$ represents the aggregated distributed power allocation.

If sensor $n$ receives the server's packet, the power allocation action $\mathbf{a}^c_{n, k}$ will be applied; otherwise, the local action $\mathbf{a}^d_{n, k}$ will be in effect. Let $\beta_{n, k} \in \{1, 0\}$ denote whether sensor $n$ receives the server's packet over the control channel at time $k$.
Thus, the collaborative distributed and centralized data channel power allocation is formulated as
\begin{equation}
\label{equation_action_comb_rule}
    \mathbf{a}_{k} =  \mathbf{a}^c_{k} \boldsymbol{\beta}_k + \mathbf{a}^d_{k} (\mathbf I - \boldsymbol{\beta}_k),
\end{equation}
where $\mathbf I$ is the $N$-by-$N$ identity matrix, and the diagonal matrix $\boldsymbol{\beta}_k \triangleq \text{diag} 
\{\beta_{1, k}, \dots, \beta_{N, k}\}$ serves as an \textbf{action collaboration indicator}, managing the blend of centralized and distributed actions determined by the control channel and attack status.

\section{Collaborative Distributed-Centralized
DRL}\label{sec:DRL}
We aim to determine the optimal collaborative power allocation action $(\mathbf a^d_k,\mathbf a^c_k)$. 
For the distributed actions, sensor $n$ generates $\mathbf a^d_{n,k}$ based on its local observation of its AoI and channel power gains:
\begin{equation}
    \mathbf{O}^{d}_{n, k}  = [\tau_{n, k-\!1}, \mathbf{g}_{n, k}^{\top}].
\end{equation}
The distributed policy at sensor $n$, $\kappa_n(\cdot)$ maps its local observation $\mathbf{O}^{d}_{n, k}$ to local action $\mathbf{a}^d_{n, k}$.

For the centralized actions, the server has the global observation
\begin{equation}
\label{equation_Ok}
\mathbf{O}^{c}_{k} = [\mathbf{O}^{d}_{1, k}, \dots, \mathbf{O}^{d}_{N, k}].
\end{equation}
The centralized policy $\mu(\cdot)$ is a mapping from the global state $\mathbf{O}^{c}_k$ to the centralized action $\mathbf{a}^c_{k}$. 

{ 
This collaborative power allocation problem is inherently a decision-making challenge, as each action depends on the current state (i.e., the observations) and directly influences future states according to~\eqref{eq:tau_update}. 
To address this, we formulate it as a long-term discounted decision-making problem:
\begin{equation}
\label{equation_overallproblem}
\begin{aligned}
& \min_{\kappa_1(\cdot), \ldots, \kappa_N(\cdot); \mu(\cdot)} \ \mathbb{E}\left[ \sum_{k=1}^{\infty}\delta^{k-1}  c_k\right] \\
\text{s.t.}\  &\mathbf{a}^d_{n, k} =  \kappa_{n}(\mathbf{O}^{d}_{n, k}),\ \mathbf{1}^\top \mathbf{a}^d_{n, k} \leq E^{s}, \forall n, \forall k \in \mathbb{Z}^{+}, \\
&\mathbf{a}^c_{k} = \mu (\mathbf{O}^{c}_k),\  \mathbf{1}^\top \mathbf{a}^c_{n, k} \leq E^{s}, \forall n, \forall k \in \mathbb{Z}^{+},
\end{aligned}
\end{equation}
where $\mathbf{1} = [1, \ldots, 1]^\top \in \mathbb{R}^M$,  $c_k$ is the per-step estimation MSE defined in~\eqref{eq:ek} and $\delta \in (0, 1)$ is the discount factor.
The objective of this optimization is to minimize the expected sum of discounted estimation MSE across all sensors, taking into account the power budget constraints $E^s$ for each sensor in both the centralized and distributed scheduling policies.}

To solve the problem, we resort to DRL, which is recognized for its effectiveness in handling complex decision-making scenarios with high-dimensional spaces. However, existing DRL methods typically employ either a centralized or distributed approach, which is not suitable for our problem that requires a collaborative design of both distributed and centralized policies.
To tackle this, we propose the collaborative distributed and centralized (CDC)-DRL framework. 

\subsection{Network Architecture of CDC-DRL}
CDC-DRL has three types of deep neural networks (DNNs), as illustrated in Fig.~\ref{fig:system}:

1) A centralized actor $\mu$ with parameter $\theta^{\mu}$ represents the central policy as
$\mathbf{a}^{c}_{k} = \mu(\mathbf{O}^c_k \vert \theta^{\mu})$;

2) $N$ distributed actors $\kappa_{1}, \ldots, \kappa_{N}$ with parameters $\theta^{\kappa_1},\dots,\theta^{\kappa_N}$, respectively, represent the distributed policy as $\mathbf{a}^{d}_{n, k} = \kappa_{n}(\mathbf{O}^{d}_{n, k} \vert \theta^{\kappa_{n}}), \forall n$;

3) A centralized critic $Q$ with parameter $\theta^{Q}$ approximates the Q function $Q(\mathbf{O}^c_k, \mathbf{a}_k \vert \theta^Q)  \approx  \mathbb{E} \left[ \sum_{k'=k}^{\infty}\delta^{k'-k}  c_{k'} \bigg \vert \mathbf{O}^c_k, \mathbf{a}_k\right]$, which represents the expected future error given the state and action under the CDC policies $\kappa_1(\cdot), \ldots, \kappa_N(\cdot)$ and $\mu(\cdot)$. This estimation will guide the actors during training, optimizing their policies to minimize expected errors, and it will not be needed for deployment.

In particular, each distributed actor’s output layer is designed with two heads to determine the sensor's transmission power. The first head outputs a scalar, which is passed through a sigmoid activation function and then scaled by the energy budget $E^s$. The second head outputs a vector of size $M$ using a softmax activation function. The product of these outputs provides the transmission power satisfying the power constraint in~\eqref{equation_overallproblem}. The centralized actor is configured in a similar manner.

\subsection{Two-stage Training of CDC-DRL}
The joint training of CDC policies is challenging, particularly due to the complex interactions between distributed and centralized decision-making under dynamic conditions. The presence of the DoS attacker with stochastic behavior further complicates these interactions, impacting training speed and stability. To address these challenges, our CDC-DRL framework adopts a two-stage training approach. Initially, separate pre-training for centralized and distributed policies optimizes the initial DNN parameters, providing a stable starting point and isolating training from adversarial disruptions. This preparation enables effective joint training, where these policies are fine-tuned to synergistically generate power allocation commands over the dynamic DoS attacks. This strategy not only quickens convergence but also improves the resilience and efficacy of the CDC-DRL algorithm.
\subsubsection{Separate Centralized and Distributed Pre-training}
In this stage,  we first train the centralized critic and actor using a classic centralized DRL approach, where $\mathbf{a}_{k} = \mathbf{a}^{c}_{k}$. The training follows an off-policy scheme based on transitions consisting of the global state, collaborative action, error, and next global state, denoted as $\langle \mathbf{O}^c,\mathbf{a},c,\widetilde{\mathbf{O}}^c \rangle$. At each time $k$, we store the transition into a replay buffer.
The optimal parameters $\theta^{Q}$ and $\theta^{\mu}$ should satisfy the Bellman equation \cite{lillicrap2019continuous} as
\begin{equation}
\label{equation_bellman}
Q(\mathbf{O}^c_k, \mu(\mathbf{O}^c_k \vert \theta^\mu) \vert \theta^{Q}) \! = \! \mathbb{E} \! \left[c_k \! + \! \delta Q(\mathbf{O}^c_{k+1}, \mu(\mathbf{O}^c_{k+1} \vert \theta^\mu) \vert \theta^{Q})\right] \!.
\end{equation}
$\theta^{Q}$ and $\theta^{\mu}$ are updated iteratively. In each iteration, $\theta^{Q}$ is first updated via gradient descent to minimize the expected temporal difference (TD) error, using a mini-batch of $\mathcal{N}$ transitions sampled from the replay buffer:
\begin{equation}
\label{equation_stage1_central_loss}
     TD^{\text{Sep}} \! = \! \frac{1}{\mathcal{N}} \sum_{i = 1}^{\mathcal{N}} \left[c_{i}  \! + \! \delta Q(\widetilde{\mathbf{O}}^c_{i}, \mu(\widetilde{\mathbf{O}}^c_{i} \vert \theta^\mu) \big \vert \theta^Q) \! - \! Q(\mathbf{O}^c_{i},\mathbf{a}_{i} \vert \theta^Q)\right]^2 \!\!\!\!.
\end{equation}
The rationale behind TD error minimization is to narrow the gap between the left-hand side and right-hand side of the Bellman equation~\eqref{equation_bellman}.
Next, $\theta^{\mu}$ is updated by gradient descent to minimize the loss $J^c$, which is the expected error estimated by the centralized critic:
\begin{equation}
\label{equation_stage1_central_actor}
J^c =  \frac{1}{\mathcal{N}} \sum_{i = 1}^{\mathcal{N}} Q(\mathbf{O}^c_{i}, \mu(\mathbf{O}^c_{i}|\theta^\mu) \big\vert \theta^Q).
\end{equation}
After training $\theta^{Q}$ and $\theta^{\mu}$, we proceed to train the distributed actors. By the guidance of centralized critic, the distributed actor DNN parameters $\theta^{\kappa_1}, \dots \theta^{\kappa_N}$ are updated by minimizing the expected error as
\begin{equation}
\label{equation_stage1_decentral_actor}
    J^d = \frac{1}{\mathcal{N}} \sum_{i = 1}^{\mathcal{N}} Q\Big(\mathbf{O}^c_i, \mathbf{a}^d_i \Big \vert \theta^Q, \mathbf{a}^d_{n, i} =  \kappa_{n}(\mathbf{O}^{d}_{n, i} \vert \theta^{\kappa_n}), \forall n \Big).
\end{equation}

\subsubsection{Joint Centralized and Distributed Training}
In contrast to the separate centralized and distributed pre-training, joint training under DoS attacks employs a collaborative power allocation action defined in~\eqref{equation_action_comb_rule}. Our goal is to train the centralized critic to estimate future errors from the global state, $\mathbf{O}^c_k$, and the executed (collaborative) action, $\mathbf{a}_k$, and use this to guide actor training.

Unlike pre-training, where the subsequent action is determined solely by the next state and the policy of either the centralized or distributed actor, in joint training, the next action becomes stochastic. This stochasticity arises from the collaborative actions of the centralized and distributed actors under the influence of the DoS attack. To enhance the training process, we introduce a new transition format, consisting of the global state, both centralized and distributed actions, the action collaboration indicator, error, next global state, and next collaborative action, denoted as $\langle \mathbf{O}^c,\mathbf{a}^c, \mathbf{a}^d, \boldsymbol{\beta}, c, \widetilde{\mathbf{O}}^c, \widetilde{\mathbf{a}}\rangle$. 

Following this, we define the critic’s loss function for the joint training stage similar to that in~\eqref{equation_stage1_central_loss}:
\begin{equation}
\label{equation_stage2_central_loss}
    TD^{\text{Joint}} = \frac{1}{\mathcal{N}} \sum_{i = 1}^{\mathcal{N}} \left[c  + \delta Q(\widetilde{\mathbf{O}}^c_{i}, \widetilde{\mathbf{a}}_{i} \big \vert \theta^Q) - Q(\mathbf{O}^c_i,\mathbf{a}_i \vert \theta^Q)\right]^2. 
\end{equation}
Subsequently, the loss function for jointly optimizing both the centralized and distributed actors is defined as:
\begin{equation}
\label{equation_stage2_actor}
J^{\text{Joint}} \! = \! \frac{1}{\mathcal{N}} \! \sum_{i = 1}^{\mathcal{N}} Q(\mathbf{O}^c_{i}, \mathbf{a}_{i} \big\vert \theta^\mu, \theta^{\kappa_1}, \dots, \theta^{\kappa_N}, \theta^Q).
\end{equation}
In \eqref{equation_stage2_central_loss} and \eqref{equation_stage2_actor}, $\mathbf{a}_{i} = \mathbf{a}^c_{i} \boldsymbol{\beta}_{i} + \mathbf{a}^d_{i} (\mathbf{I} - \boldsymbol{\beta}_{i})$ is introduced in \eqref{equation_action_comb_rule}. 
The parameters $\theta^{Q}$, $\theta^{\mu}$ and $\theta^{\kappa_1}, \dots \theta^{\kappa_N}$ are updated iteratively by gradient descent methods to minimize these loss functions.

\section{Numerical Results}
\begin{table}[t]\captionsetup{font=footnotesize}
\caption{Simulation Parameters}
\label{tab:para}
\centering
 \begin{tabular}{||c | c||} 
 \hline
 Process state transition matrix, $A_{n}$ &  Randomly generated\(^*\) \\
 Channel states, $\mathbb{G}$ & $\{10^{-4}, 10^{-3.5}, \dots, 10^{-0.5}\}$ \\
 Channel state transition matrix, $\mathbf{M}_{n}$ & \begin{tabular}[c]{@{}c@{}}Randomly generated \\ from uniform distribution\end{tabular}   \\
 Codeword length, $L$ & 200 \\
 Transmission rate, $R$ & 2 \\
 Background white noise, $\sigma^2$ &0.01 \\
 Sensors' power budget, $E^s$ & $20$ \\
 Server's power budget, $E^d$ & $200$ \\
 Attacker's power budget, $E^a$ & $1000$ \\
 \hline 
\end{tabular}
\begin{tablenotes}
\centering
\item \(^*\)Spectral radius is sampled uniformly from (1, 1.3).
\end{tablenotes}
\end{table}

We present numerical results of the CDC-DRL algorithm for transmission power allocation of remote estimation systems, based on the simulation parameters summarized in Table \ref{tab:para}. 
We consider three DoS attackers with distinct Markovian action transition matrices \(T^a \), as shown in Table \ref{tab:attacker}. While all attackers share the same expected attack frequency, their behaviors differ: the dynamic attacker frequently changes actions, the balanced attacker has moderate variability, and the persistent attacker maintains consistent attacking or non-attacking behavior.
Our results show that the persistent attacker leads to the worst performance of the defending algorithms, and we will present only this scenario due to space limitations.

\begin{table}[t]
\centering
\captionsetup{font=footnotesize}
\caption{Attacker Dynamics}
\label{tab:attacker}
\begin{tabular}{||c|c|c|c||}
\hline
Attacker & Balanced & Dynamic & Persistent \\ \hline 
& & & \\ [-2.5mm]
 $T^a$ & $\begin{bmatrix}
0.5 & 0.5\\
0.5 & 0.5 \\
\end{bmatrix}$ & $\begin{bmatrix}
0.1 & 0.9\\
0.9 & 0.1 
\end{bmatrix}$ & $\begin{bmatrix}
0.9 & 0.1\\
0.1 & 0.9 
\end{bmatrix}$ \\ [2.5mm]

\hline 
\end{tabular}
\vspace{-0.5cm}
\end{table}

The centralized critic has two hidden layers with 1024 neurons each. Each centralized or distributed actor has two hidden layers with 256 neurons. The activation function of all hidden layers is ReLU. The number of training episodes is 500, and there are 500 time-slots in each episode. Our simulation compares the performance of CDC-DRL against the following benchmarks:  

1) Centralized DDPG agent (DDPG): A purely centralized DRL approach where $\mathbf{a}^c_{k}$ is generated by DDPG~\cite{lillicrap2019continuous}. There is no sensor data channel transmission once the centralized control packet is lost, i.e., $\mathbf{a}^d_{n, k} = \mathbf{0}, \forall n$.

2) Distributed DDPG agent (MADDPG): A purely distributed DRL strategy with $\mathbf{a}^d_{n, k}$ generated by Multi-Agent DDPG\cite{Lowe2017Multi}, $\forall n$.

3) Centralized DDPG combined with distributed DDPG (DDPG+MADDPG): A straightforward combination of centralized actions from DDPG with distributed actions from MADDPG.

\begin{figure}[t]
     \centering
     \begin{subfigure}{0.22\textwidth}
         \centering
         \includegraphics[width=\textwidth]{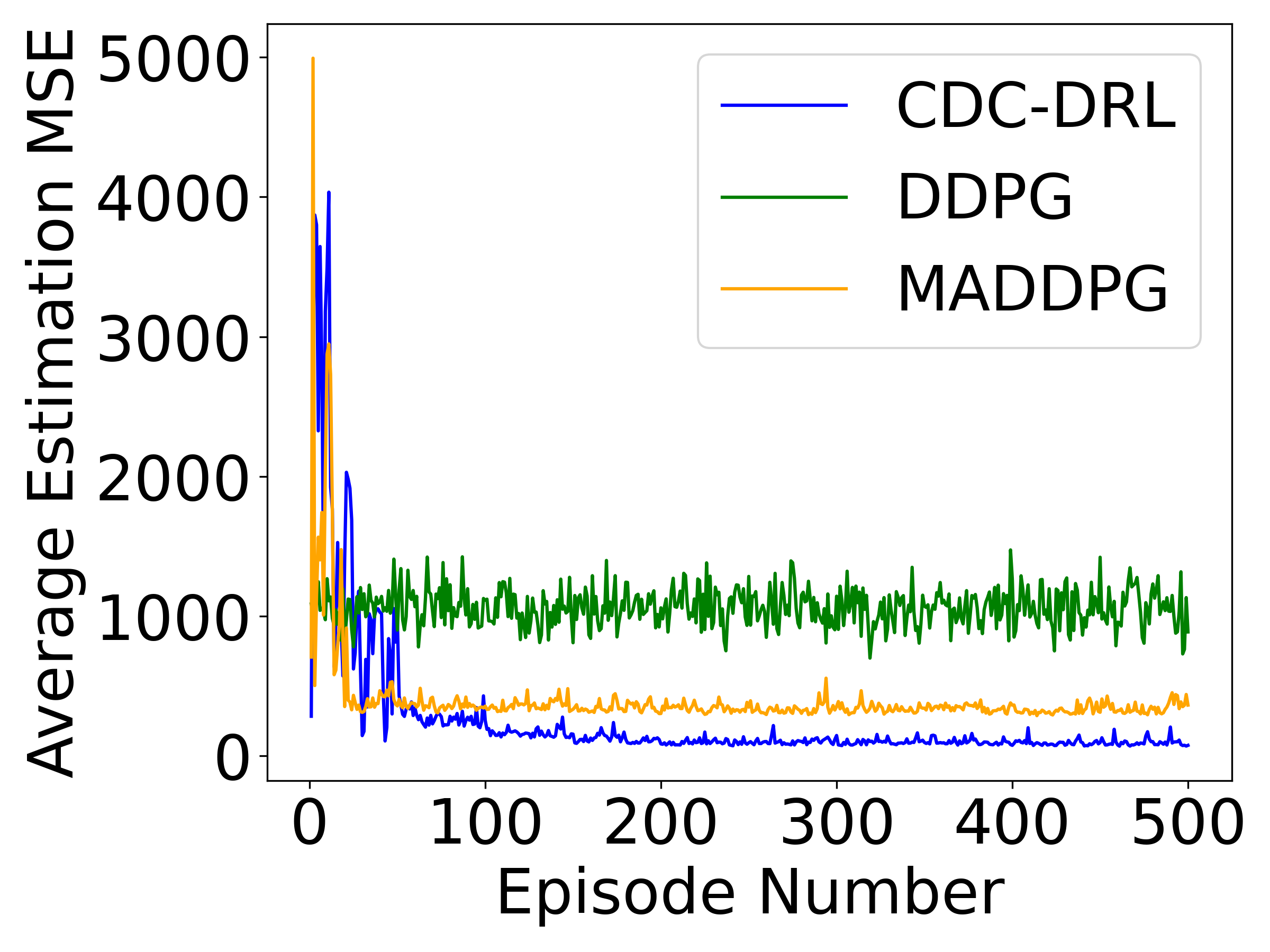}
    \captionsetup{font=footnotesize}            \vspace{-0.6cm}
         \caption{{Training processes of LSS \#1}}
         \label{subfig:train}
     \end{subfigure}
    \captionsetup{font=footnotesize}      
    \begin{subfigure}{0.22\textwidth}
         \centering
         \includegraphics[width=\textwidth]{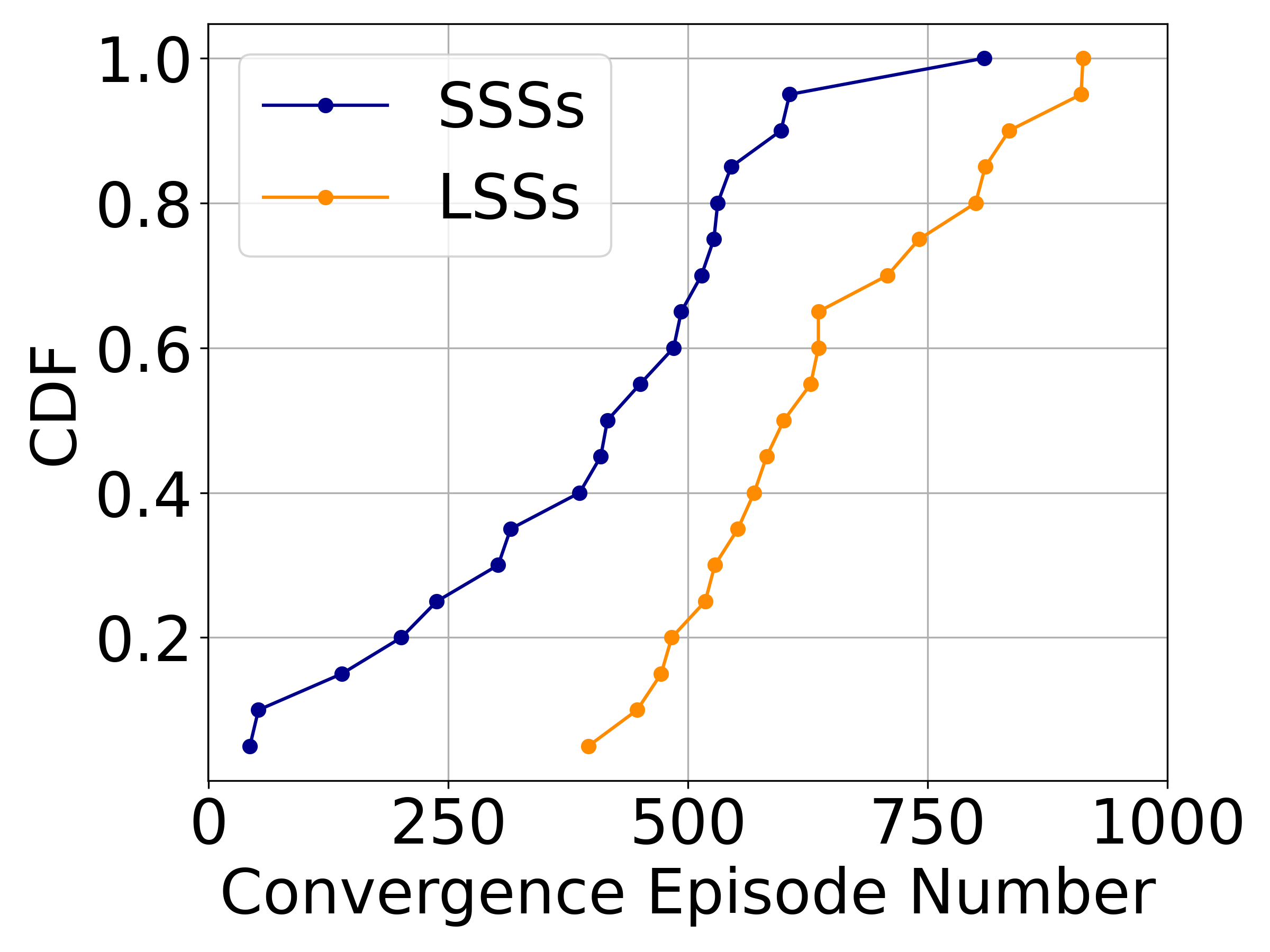}
    \captionsetup{font=footnotesize}             \vspace{-0.6cm} 
         \caption{{CDF of convergence time}}
         \label{subfig:cdf}
     \end{subfigure}
\caption{{Convergence analysis of CDC-DRL}}
\label{fig:converge}
\vspace{-0.5cm}
\end{figure}

{ The performance of CDC-DRL in terms of the average estimation MSE is illustrated in Fig.~\ref{fig:results}, where it is compared against benchmarks in both small-scale systems (SSSs) with 6 sensors and 3 channels (Fig.~\ref{subfig:small}) and large-scale systems (LSSs) with 20 sensors and 10 channels (Fig.~\ref{subfig:large}) across 20 different system setups. System parameters are randomly generated based on Table~\ref{tab:para} and identified by System ID. 
We see that CDC-DRL consistently outperforms all benchmarks, achieving a notably lower average error and more stable performance across different systems, and the conventional centralized approach (DDPG) is the worst impacted by DoS attacks, especially in LSSs.
In particular, CDC-DRL achieves an average reduction in average estimation MSE of 13.0\% compared to the best benchmarks across all systems in the SSS setting, and 52.6\% in the LSS setting.
This shows that the proposed algorithm effectively addresses the problem in LSSs with complex, high-dimensional data, delivering substantial improvements in estimation accuracy and scalability.

\begin{figure}[t]
     \centering
     \begin{subfigure}{0.48\textwidth}
         \centering
         \includegraphics[width=\textwidth]{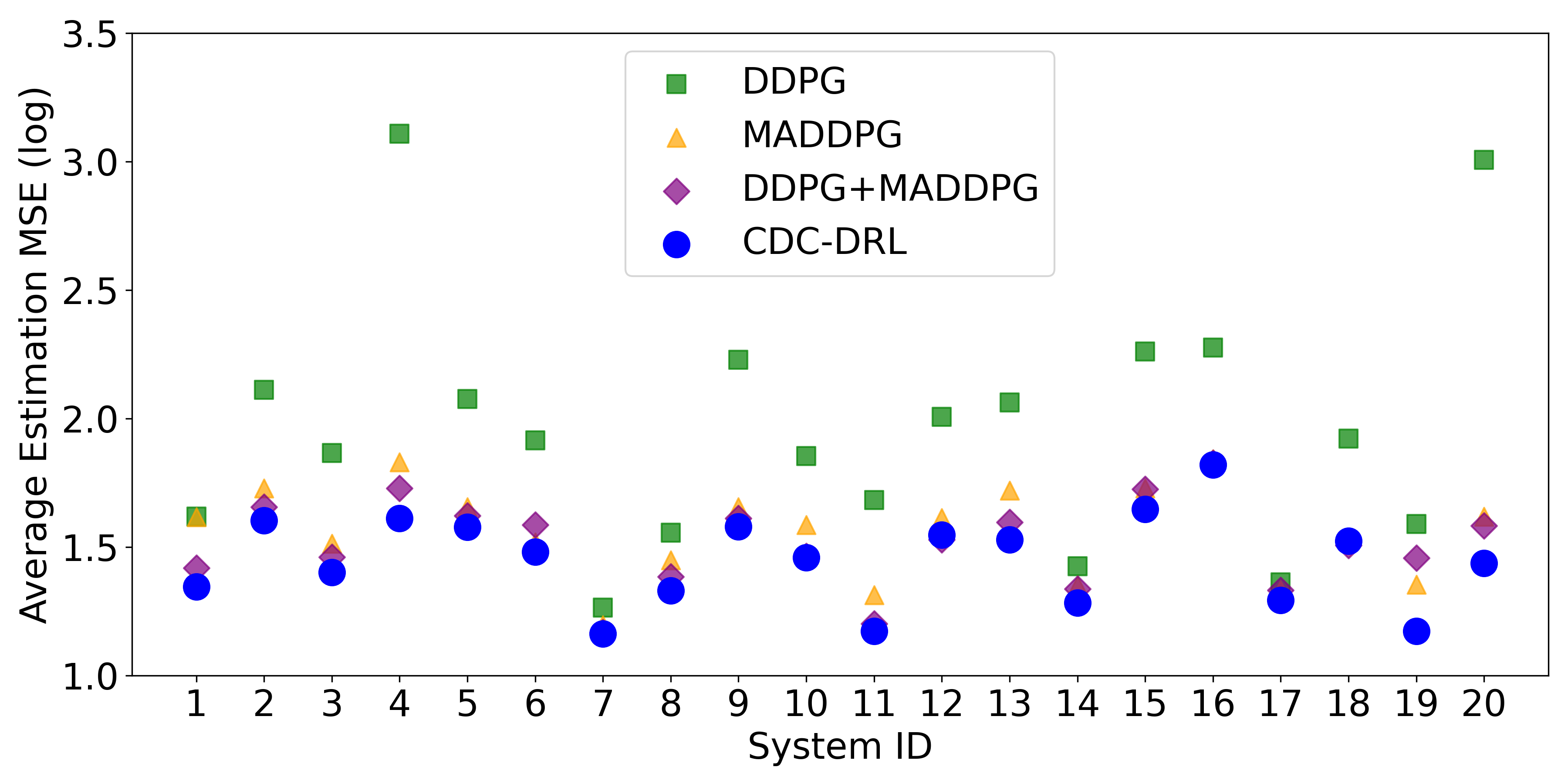}
    \captionsetup{font=footnotesize}             \vspace{-0.6cm} 
         \caption{{ Small-scale systems (SSSs): $N=6$ and $M=3$}}
         \label{subfig:small}
     \end{subfigure}
     \hfill
     \begin{subfigure}{0.48\textwidth}
         \centering
         \includegraphics[width=\textwidth]{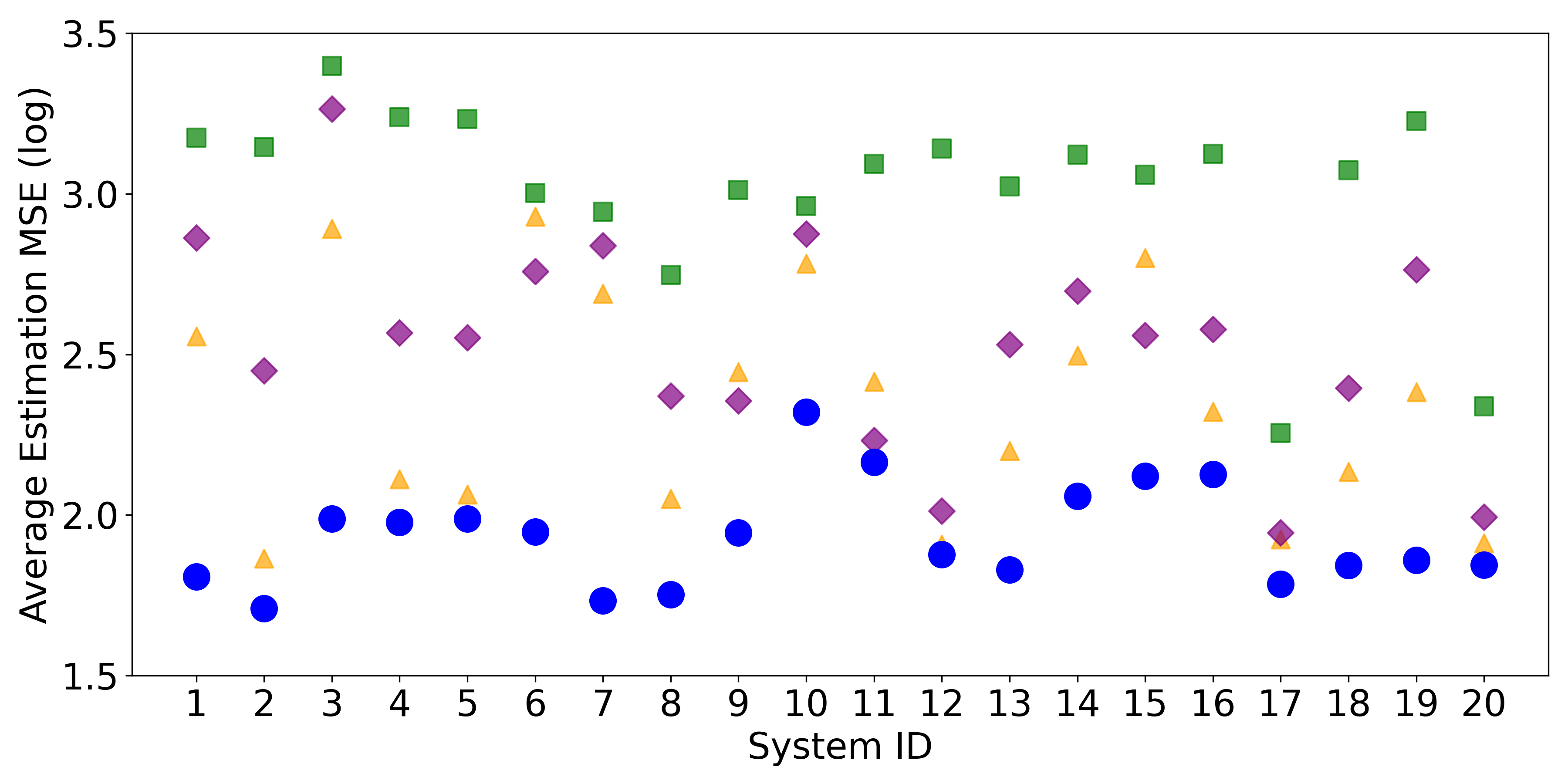}
    \captionsetup{font=footnotesize}             \vspace{-0.6cm}
         \caption{{ Large-scale systems (LSSs): $N=20$ and $M=10$}}
         \label{subfig:large}
     \end{subfigure}
    \captionsetup{font=footnotesize}    
\caption{{ Performance comparison between CDC-DRL and the benchmarks.}}
\vspace{-0.5cm}
\label{fig:results}
\end{figure}

An empirical convergence analysis of CDC-DRL is presented in Fig. \ref{fig:converge}. 
In particular, we evaluate algorithm convergence by comparing the Normalized Standard Deviation (N-std) of the estimation error’s moving average over 10 episodes, with a 5\% threshold. Convergence is achieved if the N-std is below this threshold.
The training processes of CDC-DRL and the benchmarks with LSS \#1 from the experiment in Fig.~\ref{subfig:large} is depicted in Fig. \ref{subfig:train}. We observe that all algorithms achieve convergence, with CDC-DRL taking slightly longer to converge due to its increased complexity, but resulting in a substantial reduction in estimation MSE.
The empirical cumulative distribution function (CDF) of CDC-DRL's convergence episodes from the experiments in Fig.~\ref{fig:results} is shown in Fig. \ref{subfig:cdf}, indicating that the LSS setup clearly requires longer convergence times. On average, CDC-DRL converges in approximately 403 episodes for SSSs and 638 episodes for LSSs. Our simulations use an RTX 3070 GPU and an Intel i7-11700F CPU, with CDC-DRL training for 600 episodes in an LSS setup taking about 150 minutes. A high-performance edge server with 10x RTX 3090 GPUs could reduce this time to around 8 minutes.}
\vspace{-0.3cm}
\section{Conclusions}
We have developed a CDC-DRL framework to address the power allocation challenges in remote estimation systems under control channel DoS attacks. Our proposed solution significantly outperforms benchmarks based on either centralized or distributed DRL approaches, particularly in large-scale systems. { In future work, we will explore more complex system setups that incorporate multiple control channels and account for inaccurate channel state information.}

    \balance
\ifCLASSOPTIONcaptionsoff
\newpage
\fi


\end{document}